\documentstyle[12pt]{article}
\input psfig.tex
\begin{document}
\voffset =-0.4in

\def \beq {\begin{equation}}
\def \eeq {\end{equation}}
\def \beqn {\begin{eqnarray}}
\def \eeqn {\end{eqnarray}}
\centerline{\bf $Z'$ in the 3-3-1 model}

\bigskip
\centerline{Pankaj Jain and Satish D. Joglekar}

\medskip
\centerline{Physics Department}
\centerline{I.I.T.}
\centerline{Kanpur 208016}
\centerline{India}
\centerline {e-mail:pkjain@iitk.ernet.in}




\vspace{1.0cm}

{\centerline {\bf Abstract}}

\noindent
Phenomenological implications of the $Z'$ in
 SU(3)$\times$SU(3)$\times$U(1) extension of the standard
 model are studied. 
The model improves the fit for $R_b$
as well as the high $E_T$ jet cross section observed by CDF.

\vspace{1.0cm}
\noindent
{\bf 1. Model:}  We consider the SU(3)$_{\rm color}\times$SU(3)$\times$U(1)
extension of the standard model, proposed earlier by Frampton [1] 
to determine its predictions for the observables that appear to
show deviation from standard model. 
The SU(3)$\times$U(1) interaction breaks
down to SU(2)$\times$U(1) at the 
scale of about 1 TeV.
The model has several interesting features.
Leptons in this model interact only with the weak SU(3), namely they
do not directly couple to U(1). The U(1)
gauge interaction with relatively strong coupling leads 
to a new leptophobic interaction. The U(1) coupling has to be
relatively strong, else it leads to a light leptophobic $Z'$ which might
already be ruled out by the CDF jet data.
The anomaly cancellation occurs among the three generations
and thus might provide a hint for their existence. 
The model also leads to doubly charged gauge bosons 
which provide lepton number violating interaction. The experimental
limit on the masses of these gauge bosons is found to be 230 GeV [2].

In the present paper we show that a slight modification of the original
model [1], leads to successful explanation
of $R_b$ as well as of the anomalous increase in the jet
cross section seen in CDF [3].
The original large discrepancy in $R_b$ [4] 
which generated considerable theoretical
interest [5-10] has dissapeared and it now
appears to be only a $2\sigma$ effect. The present model brings the
theoretical prediction further close to experiments without 
significantly effecting the total $Z$ width. It also provides a
new leptophobic $Z'$ which might be necessary to explain the large
deviation in jet cross section seen by CDF [3].
  
The lepton triplets,
$$\left(\matrix{e\cr \nu_e\cr L_e}\right)_L\ \ ,\ \ \left(\matrix{\mu\cr \nu_\mu
\cr L_\mu}\right)_L\ \ ,\ \ \left(\matrix{\tau\cr \nu_\tau\cr L_\tau}\right)_L
$$
are assigned to the $\bar 3$ representation of the SU(3) group. The
leptons $L_e$, $L_\mu$ and $L_\tau$ can be identified with $e^c_L$,
$\mu^c_L$, $\tau^c_L$. The right handed neutrinos are singlets.
The quark triplets 
$$
\left(\matrix{c\cr s\cr q_c}\right)_L\ \ ,\ \ 
\left(\matrix{t\cr b\cr q_t}\right)_L\ \ ,\ \ 
$$
are assigned to the $3$ representation and have strong hypercharge 
$Y_S = 2/3$ and the triplet 
$$ \left(\matrix{d\cr u\cr q_u}\right)_L $$
to the $\bar 3$ representation and has strong hypercharge $Y_S = -4/3$. 
Note that this assignment is reveresed in comparison to what was
proposed in Ref. [1]. This is necessary so that the correction to $R_b$
is positive in this model.

It is not possible to identify the third members of the 
quark triplets with one of the known right handed quarks since the hypercharges
do not agree. These have to be identified with new exotic heavy quarks.
The right handed quarks are all singlets under SU(3) with strong hypercharge
assignments given by, 
$$Y_S(u_R) = Y_S(c_R) = Y_S(t_R) = -4/3$$
$$Y_S(d_R) = Y_S(s_R) = Y_S(b_R) = 2/3$$
$$Y_S(q_{cR}) = Y_S(q_{tR}) = 8/3 $$
$$Y_S(q_{uR}) = -10/3$$
The anomalies cancel with this assignment. 

The Higgs fields necessary for  breakdown of this symmetry 
are discussed in section 3. We 
introduce three higgs fields $\phi$, $\phi_1$, and $\phi_2$ which
respectively form 
$(3,-2)$, $(3,0)$ and $(3,2)$
representations of the SU(3)$\times$ U(1). 
The vacuum values of these fields have
been assumed to be
\beq<\phi> = {1\over \sqrt {2}}\left(\matrix{0\cr 0\cr v\cr}\right)\ ,\
 <\phi_1> = {1\over \sqrt {2}}\left(\matrix{0\cr v_1\cr 0\cr}\right)\ ,\ 
<\phi_2> = {1\over \sqrt {2}}\left(\matrix{v_2\cr 0\cr 0\cr}\right)\eeq
We discuss the Higgs potential that can yield such a vev and also
the resulting physical Higgs masses in next section.
We will assume that $v>>v_1,\ v_2$ and is responsible for the breakdown
of SU(3)$\times$U(1) down to SU(2)$\times$U(1). The scale of this
symmetry breaking is determined
by fitting the experimental value of $\sin^2\theta_W$.
In the present
model $\sin^2\theta_W$ at the symmetry breaking scale is found to be
\beq \sin^2\theta_W = {1\over 4[1+g^2/(4g_S^2)]}
\eeq 
$g$ and $g_S$ are the SU(3) and U(1) coupling constants respectively 
 at the symmetry breaking scale. The SU(3) gauge bosons corresponding
to the broken generators will be denoted as $Y^{++}$, $Y^{--}$, $Y^+$
and $Y^-$, consistent with the notation used in Ref. [1] and will
acquire masses $M(Y) \approx gv/2$. We require that the $\sin^2\theta_W$
measured at $M_Z$ in the $\overline{\rm MS}$ scheme and evolved to energy 
scale $M(Y)$ should agree with the
value given in Eq. [2]. The resulting values of $g_S$ and $v$ are displayed
in Table 1. The corresponding values of $M(Y)$ and $M(Z')$, where
$Z'$ is the vector boson corresponding to the superposition of
the U(1) and $T_8$ generator of SU(3), are also shown. $Z'$ effectively
acts as a leptophobic gauge boson and could lead to the enhancement
in jet cross section at large $E_T$ needed to explain the CDF data.
We discuss this later.
The masses of $W^\pm$ in the present model turn out to be
$M(W^\pm) = g\sqrt{v_1^2+v_2^2}/2$. 
The electroweak hypercharges are given by 
$$ Y = \sqrt 3\lambda_8 - Y_S$$

\bigskip
\begin{center}
\begin{tabular}{|c|c|c|c|}
\hline
M(Z') & $g_S$ & M(Y) & v\\
GeV & & GeV & GeV\\
\hline
1000 & 1.47 & 221 & 680\\
\hline
1200 & 1.54 & 254 & 779\\
\hline
1400 & 1.61 & 283 & 870\\
\hline
1600 & 1.66 & 314 & 964 \\
\hline
1800 & 1.70 & 345 & 1059\\
\hline
2000 & 1.79 & 364 & 1117\\
\hline
\end{tabular}
\end{center}
\bigskip
\noindent
Table 1: The masses of the gauge bosons $Z'$, $Y$ as a function of the 
U(1) gauge coupling $g_s$. The corresponding 
 vacuum expectation values $v$ of the higgs
field $\phi$ are also shown.

\bigskip
\noindent
{\bf 2. Results:}  
We next determine the values of $R_b$ and $R_c$ within the present model.
The model leads to corrections of the order $(v_1/v)^2$ 
and $(v_2/v)^2$
beyond the standard model. These corrections  
 arise from the mixing of the electroweak $Z$ boson with the 
leptophobic $Z'$. The symmetry breakdown occurs in two stages,
first SU(3)$\times$U(1) breaks to SU(2)$\times$U(1) at the scale
$v\approx 1$ TeV, which then further breaks down to U(1) at the 
electroweak scale. We identify the three neutral physical bosons
$Z$, $Z'$, and the photon approximately
to order $v_1^2/v^2$ and $v_2^2/v^2$. 
The $Z$ and $Z'$ couplings to the fermions turn out to be,
\beqn
{\cal L}_{Z,Z'} &=& \bar\psi_L\left[{g\over \cos\theta_W}
\left(-Q\sin^2\theta_W + T_3\right)
+ {\delta\over 2}g_SY_S^L\right]\gamma_\mu Z^\mu\psi_L\nonumber\\ 
& + &
\bar\psi_R\left[-{g\over \cos\theta_W}\sin^2\theta_W Q + 
{\delta\over 2}g_SY_S^R\right]\gamma_\mu Z^\mu\psi_R \nonumber\\ 
&+& \bar\psi_L g_S {Y_S^L\over 2}\gamma_\mu Z'^\mu\psi_L
+ \bar\psi_R g_S {Y_S^R\over 2}\gamma_\mu Z'^\mu\psi_R
\eeqn
Here we have neglected terms proportional to $g^2\tan^2\theta_W$ in 
comparison to $g_S^2$ since they give very small corrections.
We have also set $\cos\theta= g_S/\sqrt{g'^2+g_S^2} \approx 1$, where 
$g' = g/\sqrt{3}$ at the scale of SU(3) breaking.
We find that the mixing angle between
$Z$ and $Z'$ is given by,
$$\delta \approx {g\over 2g_S^2v^2\cos\theta_W}\left[{g\tan^2\theta_W
 (v_2^2
-v_1^2)\over 2 g_S} - g_S\ v_2^2\right]$$
The $Z$ width is then found to be proportional to 
$$(t_3-4Q\sin^2\theta_W)^2 + (t_3)^2 + {4\cos\theta_W\over g} t_3g_SY_S^L\delta
- {8\cos\theta_W\over g} Q \sin^2\theta_W g_S(Y_S^L+Y_S^R)\delta$$
where $t_3$ is $+1$ for up type quarks and $-1$ for down type 
quarks and $Q$ is
the electric charge. In this formula we have neglected 
 terms of order $(g\tan\theta_W)^2$ which are an order of
magnitude smaller in comparison to $g_S^2$.
We find that the resulting correction to leptonic $Z$ width is very small.
All the hadronic decay modes, however, receive a correction. The total
hadronic decay width receives a very small correction which is smaller than
the current uncertainty in its measured value.
The resulting corrections to $R_b$, $R_c$ and the total hadronic width are
shown in table 2 for several different values of $M_Z'$ and $v_2$. We point
out that $v_2$ is so far an adjustable parameter.

\bigskip
\begin{center}
\begin{tabular}{|c|c|c|c|c|}
\hline
$M_Z'$ & $v_2$ & $\delta R_b/R_b$ & $\delta R_c/R_c$ & $\delta \Gamma_h/\Gamma_h$ \\
\hline 
1000 & 30 & 0.0047 & $-$0.0075 & $-$0.0011\\
\hline
1000 & 40 & 0.0058 & $-$0.0094 & $-$0.0014\\
\hline
1100 & 40 & 0.0049 & $-$0.079 & $-$0.0012\\ 
\hline
1100 & 50 & 0.0062 & $-$0.010 & $-$0.0015\\
\hline
1200 & 40 & 0.0043 & $-$0.0068 & $-$0.0010\\
\hline
1200 & 50 & 0.0054 & $-$0.0087 & $-$0.0013 \\
\hline
1400 & 50 & 0.0042 & $-$0.067 & $-$0.0010\\
\hline
1400 & 60 & 0.0053 &$-$0.0086&$-$0.0013 \\ 
\hline
1600 & 60 & 0.0043 & $-$0.0069 & $-$0.0010\\
\hline
1600 & 70 & 0.0054 & $-$0.0086 & $-$0.0013\\
\hline
1800& 70 & 0.0044 & $-$0.0071 & $-$0.0011\\
\hline
1800 & 80 & 0.0055 & $-$0.0088 & $-$0.0013\\
\hline
2000 & 80 & 0.0048 & $-$0.0078 & $-$0.0012\\
\hline 
2000 & 90 & 0.0059 & $-$ 0.0095 & $-$0.0014\\
\hline
\end{tabular}
\end{center}
\bigskip
Table 2: The shift in $R_b$, $R_c$ and hadronic width due to mixing
with $Z'$. Several representative cases of $Z'$ mass for different 
values of $v_2$ are shown. $v_2$ can be treated as an adjustable
parameter to obtain the best experimental fit. 
For comparison $R_b^{\rm exp} = 0.2179\pm 0.0012$ [11], 
$R_b^{\rm SM} = 0.2158$, 
$R_c^{\rm exp} = 0.1715\pm 0.0056$ [11], $R_c^{\rm SM} = 0.1723$. 
$\Gamma_h^{\rm exp} = 1.7448\pm 0.030$ GeV 
and $\Gamma_h^{\rm SM} = 1.746$ GeV.

\bigskip
In table 2. we have shown some typical results for several values
of the mass of $Z'$ and the vacuum expectation value $v_2$. The 
model reduces the observed discrepancy  
in $R_b$. The experimental results are $R_b^{\rm exp} 
= 0.2179\pm 0.0012$ [11], $R_c^{\rm exp} = 0.1715\pm 0.0056$ [11] and
$\Gamma_h^{\rm exp} = 1.7448\pm 0.030 $ GeV
 and the Standard model values are
$R_b^{\rm SM} = 0.2158$, $R_c^{\rm SM} = 0.1723$ and $\Gamma_h^{\rm SM}=1.746$
GeV.
The values of $R_b$ and $R_c$ in the present model are both
within one sigma of the experimental results with the range of 
parameters shown in Table 2. 
The model produces very small change in the total hadronic width. 
The shift in other observables such as the 
asymetries is very small compared to the experimental error in these
quantities. The resulting correction to the jet cross section is 
shown in Fig. 1 for three different values of $M_Z'$. It is clear
from the figure that for certain range of parameters
the model can explain the deviation observed in jet cross section 
by CDF. In all our discussion
we have concentrated on somewhat large values of $g_S \ge 1.47$. For
smaller values the model produces a large correction to the jet cross section
and might already be ruled out as is clear from Fig. 1.  

\psfig{file=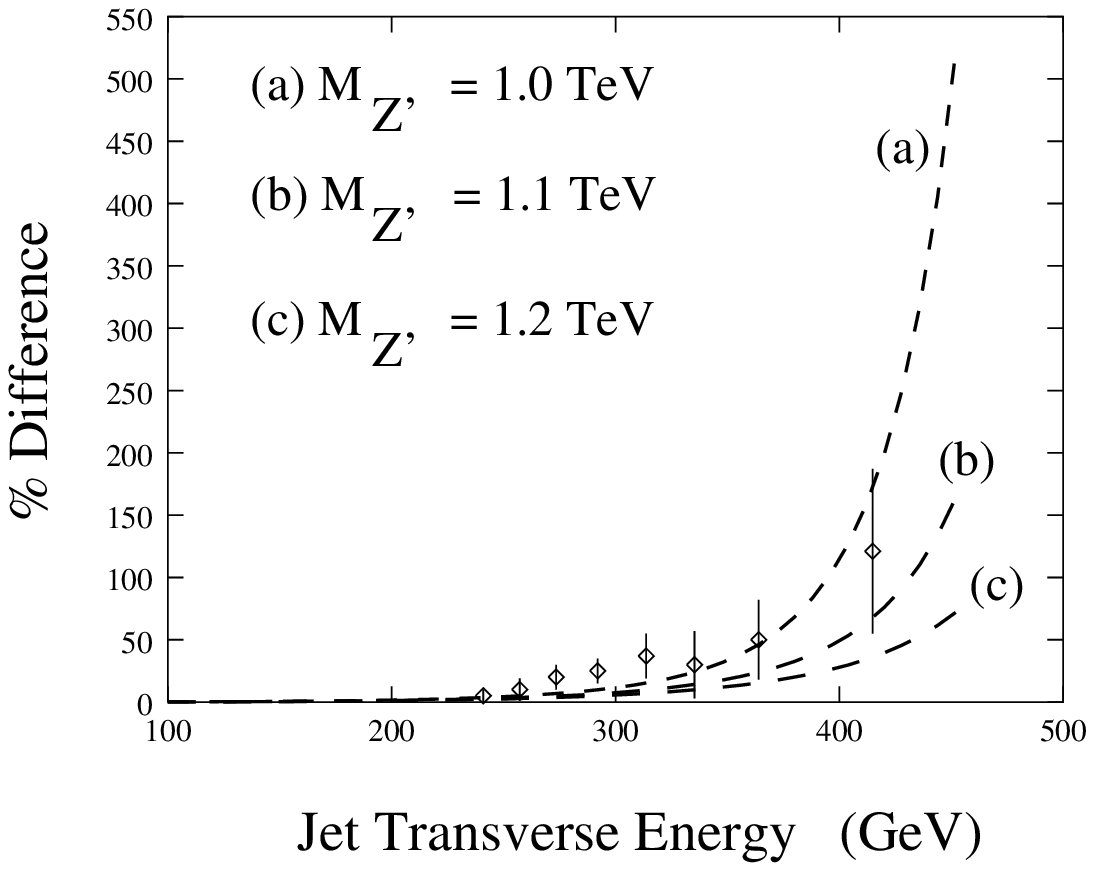}

{\noindent {\bf Fig. 1:} The \% shift due to $Z'$ in the 
jet cross section at large transverse momentum in comparison to the 
deviation from the standard model observed by CDF. The three chosen 
$Z'$ mass values 1.0 TeV, 1.1 TeV and 1.2 TeV correspond to the
U(1) coupling $g_S$ values of 1.47, 1.5 and 1.54 respectively}

\vglue 0.5in
\noindent
{\bf 3. Higgs Potential:}
In the analysis so far we have introduced three higgs fields
$\phi$, $\phi_1$, $\phi_2$ which are in the representation
$$ \phi:\ (3,-2)\ ,\ \phi_1:\ (3,0)\ ,\ \phi_2:\ (3,2) $$
of SU(3)$\times$U(1). 
The Higgs potential has to be chosen such that it breaks
SU(3)$\times$U(1) to SU(2)$\times$U(1). In our analysis we
have assumed that the symmetry breaking occurs such that 
the three higgs fields acquire vacuum expectation values 
given in equation 1. 
The Higgs potential that can accomplish this symmetry breaking 
can be chosen to have the form,
\beqn
 V &=& -\mu^2\phi^\dagger\phi + \lambda(\phi^\dagger\phi)^2 
-\mu_1^2\phi_1^\dagger\phi_1 + \lambda_1(\phi_1^\dagger\phi_1)^2\nonumber\\
&-&\mu_2^2\phi_2^\dagger\phi_2 + \lambda_2(\phi_2^\dagger\phi_2)^2 +
\lambda_3|\phi^\dagger\phi_1|^2 + \lambda_4|\phi^\dagger\phi_2|^2\nonumber\\
&+& \lambda_5|\phi_1^\dagger\phi_2|^2 + (\lambda_6\epsilon_{ijk}\phi_{2i}\phi_{ij}
\phi_k + \lambda_7(\phi_2^\dagger\phi_1)(\phi^\dagger\phi_1) + h.c)
\eeqn
The $\lambda_6$ term assures that there are no more Goldstone particles
in the spectrum other than the one's needed to give masses to the 
gauge bosons. If $\lambda_6$ is set equal to zero then we get one extra
massless scalar particle in the physical spectrum. This is because
in the absence of the this term, the U(1) transformation 
$$\phi \rightarrow e^{i\alpha}\phi\ \ ,\ \ \phi_1 \rightarrow e^{i\alpha}
\phi_1\ \ ,\ \ \phi_2 \rightarrow e^{i\alpha}\phi_2$$
is a symmetry of the action but not of the vacuum. The $\lambda_6$ term is 
needed to explicitly break this symmetry.
 The $\lambda_7$ term has been included to break the U(1) generator $W$ 
corresponding to the transformation,
$$\phi \rightarrow e^{i\alpha}\phi\ \ ,\ \ \phi_1 \rightarrow e^{-2i\alpha}
\phi_1\ \ ,\ \ \phi_2 \rightarrow e^{i\alpha}\phi_2$$
which is a symmetry of the action in the absence of this term. In this
case, however, setting $\lambda_7=0$ does not lead to a Goldstone boson
since the vacuum remains invariant under an additional U(1) whose generator
$Q'$ is given by
$$ Q' = -{3\over 2}Y_S + 2\sqrt{3}\lambda_8 + W $$

We need to show that for a certain acceptable range of parameters this
potential does have the assumed global minima.
We may parametrize the higgs minimum as 
$$<\phi^\dagger\phi> = v^2\ \ <\phi_1^\dagger\phi_1>=v_1^2\ \ 
<\phi_2^\dagger\phi_2> = v_2^2$$
$$<\phi^\dagger\phi_1> = vv_1\sin \beta_1e^{i\psi_1}\ \ 
<\phi^\dagger\phi_2> = vv_2\sin\beta_2e^{i\psi_2}\ \ 
<\phi_1^\dagger\phi_2> = v_1v_2\sin\beta_{12}e^{i\psi_{12}}\ .$$ 
 
Minizmizing the resulting potential with respect to $\psi_1-\psi_2$ 
gives $\lambda_7\cos(\psi_1-\psi_2) = -|\lambda_7|$. Henceforth we
treat $\lambda_7>0$ and take $\cos(\psi_1-\psi_2)=-1$. We first also
set $\lambda_6=0$. 
The extremization with respect to $\beta_1$, $\beta_2$ and $\beta_{12}$
leads to the following three types of minima, assuming $v$, $v_1$ and $v_2$
are all nonzero,
\begin{itemize}
\item[I.] The three $<\phi>$s are orthogonal to one another.
\item[II.] $<\phi>$ and $<\phi_2>$ and parallel to one another but
both of them are orthogonal to $<\phi_1>$.
\item[III.] All of $<\phi>$s are parallel to one another. 
\end{itemize}

The potential at these minima is,
$$V|_{\rm min} = -{1\over 2}(\mu^2v^2 + \mu_1^2v_1^2+\mu_2^2v_2^2)\ .$$
We want $$V_{II} - V_I = -{1\over 2}(\mu^2\delta v^2 + \mu_1^2\delta v_1^2
+ \mu_2^2\delta v_2^2) > 0\ ,$$ 
and similarly $V_{III} - V_I > 0$.  
A sufficient, but not necessary, condition that will ensure this is
for each $\delta v^2_i$ to be separately negative. The $\delta v_i^2$ for 
$V_{II} - V_I$ and $V_{III} - V_I$ are negative if,
$$\lambda_3 v_1^2 + \lambda_4 v_2^2 - \lambda_7{v_1^2v_2\over v} > 0$$
$$\lambda_3 v^2 + \lambda_5 v_2^2 - 2\lambda_7 v v_2 > 0$$
$$\lambda_4v^2 + \lambda_5 v_1^2 - \lambda_7{vv_1^2\over v_2} > 0$$
Note that $v$, $v_1$ and $v_2$ are all positive numbers. The above 
conditions are complicated since 
$v_i$ are themselves determined in terms of $\lambda$'s.
If we set $\lambda_7=0$ and take $\lambda_3$, $\lambda_4$ and
$\lambda_5$ to be positive these conditions are clearly satisfied.
It is clear that, if $\lambda_7$ is nonzero, 
 there will be a range of $\lambda_7$ for which the 
above mentioned sufficient conditions are fulfilled. 
Now suppose we have a $\lambda_7$
such that $V_{II}$ and $V_{III}$ are higher than $V_I$ by atleast $\delta$.
Then there is atleast a range of $\lambda_6$ for which
$V_I$ is a global minimum. Actually $\lambda_6$ term for $\lambda_6<0$, 
gives negative contribution to $V_I$ and zero contribution to $V_{II}$
 and $V_{III}$.
Hence for $\lambda_6$ negative $V_I$ will be a global minima.

There exist a total of ten massive scalar fields in the physical 
spectrum. These are identified by expanding about the minimum of the 
potential such that
$$\phi = <\phi> + \hat\phi\ \ .$$
 We give the results only for $\lambda_7 = 0$. 
We note that none of
the massive scalar fields become massless when $\lambda_7$ is set equal
to zero. 
Setting $\lambda_7\ne 0$ therefore 
only changes the mass splittings and not the masses themselves as long
as $\lambda_7$ is not too large. 
We define
three complex scalar fields $H_1$, $H_2$ and $H_3$, 
 $$ H_1 = {v_1\hat\phi_2^*+v\hat\phi_{13}\over \sqrt{v_1^2+v^2}}\ \ ,\ \ 
H_2 = {v\hat\phi_{23}+v_2\hat\phi_1^*\over \sqrt{v_1^2+v^2}}\ \ ,\ \ 
H_3 = {v_1
\hat\phi_{22}+v_2\hat\phi_{11}^*\over \sqrt{v_1^2+v_2^2}}$$
which have masses $$M_1^2 = {(v^2+v_1^2)\over 2}
\left[\lambda_3-{\sqrt{2}\lambda_6v_2\over vv_1}\right]$$
$$M_2^2={(v^2+v_2^2)\over 2}\left[\lambda_4-{\sqrt{2}\lambda_6v_1\over vv_2}
\right]$$ 
and
 $$M_3^2={(v_1^2+v_2^2)\over 2}\left[\lambda_5-{\sqrt{2}
\lambda_6 v\over v_1v_2}\right]$$
respectively. We also have the real field 
$$N\left[{{\rm Im}\ \hat\phi_3\over v} + {{\rm Im}\ \hat\phi_{12}\over v_1}
 + {{\rm Im}
\ \hat\phi_{21}\over v_2}\right]$$
with mass $$ -{\lambda_6 v v_1 v_2\over \sqrt{2}}\left[{1\over v^2}
+ {1\over v_1^2} + {1\over v_2^2}\right].$$
Finally we have the remaining three massive real fields Re$\hat\phi_3$,
Re$\hat\phi_{12}$ and Re$\hat\phi_{21}$ with mass matrix,
$$\left(\matrix{2\lambda v^2 -\lambda_6 v_1 v_2/\sqrt{2}v &
 \lambda_6 v_2/ \sqrt{2} & \lambda_6 v_1/ \sqrt{2}\cr 
\lambda_6 v_2/ \sqrt{2} & 2\lambda_1v_1^2 - \lambda_6 v v_2
/ \sqrt{2} v_1 & \lambda_6 v/ \sqrt{2}\cr
\lambda_6 v_1/ \sqrt{2} &
\lambda_6 v/ \sqrt{2} &
2\lambda_2v_2^2 - \lambda_6 v_1v/ \sqrt{2} v_2} \right)
$$

\bigskip
\noindent
{\bf Acknowledgements:} We would like to thank John Ralston for making 
useful suggestions in this work.

\bigskip
\centerline{\bf References}
\begin{itemize}
\item[1.] P. H. Frampton, Phys. Rev. Lett. {\bf 69}, 2889 (1992).  
\item[2.] E. D. Carlson and P. H. Frampton, Phys. Lett. B {\bf 283},
123 (1992). 
\item[3.] F. Abe et al., Phys. Rev. Lett. {\bf 77} (1996) 438.
\item[4.] The LEP Collaboration, Aleph, Delphi, L3, Opal, and the 
LEP Electroweak working Group, CERN-PPE/95-172.
\item[5.] G. Alterelli, N. Di Bartolomeo, F. Feruglio, R. Gatto and
M. L. Mangano, Phys. Lett. {\bf B375}, 292 (1996);    .
\item[6.] P. Chiappetta, J. Layssac, F.M. Renard and C.
Verzegnassi, Phys. Rev. {\bf D54}, 789 (1996).
\item[7.] V. Barger, K. Cheung and P. Langacker, preprint MADPH-96-936,
UPR-0696T, UTEXAS-HEP-96-2, DOE-ER-40757-078.
\item[8.] J. Rosner, preprint hep-ph/9607207, CERN-TH/96-169,
EFI-96-25.
\item[9.] Jorge L. Lopez and D. V. Nanopoulous, 
preprint hep-ph/9605359, DOE/ER/40717-27, CTP-TAMM-20/96, ACT-06/96.
\item[10.] P. H. Frampton, M. B. Wise and B. D. Wright, preprint
IFP-722-UNC, CALT-68-2048. 
\item[11.] LEP Electroweak Working Group, LEPEWWG/96-02 (30 July 1996).
\end{itemize}

\end{document}